\documentclass[english,pra,superscriptaddress,showpacs,twocolumn]{revtex4}
\usepackage{mathrsfs}
\usepackage{amsmath}
\usepackage{amssymb}
\usepackage{graphicx}
\usepackage{esint}
\usepackage{times}

\begin{document}

\title{Cavity polariton in a quasi-lattice of qubits and its selective radiation}

\author{Hou Ian}

\affiliation{Faculty of Science and Technology, University of Macau, Macau}

\author{Yu-xi Liu}

\affiliation{Institute of Microelectronics, Tsinghua University, Beijing 100084,
China}

\affiliation{Tsinghua National Laboratory for Information Science and Technology
(TNList), Tsinghua University, Beijing 100084, China}
\begin{abstract}
In a circuit quantum eletrodynamic system, a chain of $N$ qubits
inhomogeneously coupled to a cavity field forms a mesoscopic quasi-lattice,
which is characterized by its degree of deformation from a normal
lattice. This deformation is a function of the relative spacing, that
is the ratio of the qubit spacing to the cavity wavelength. A polariton
mode arise in the quasi-lattice as the dressed mode of the lattice
excitation by the cavity photon. We show that the transition probability
of the polariton mode is either enhanced or decreased compared to
that of a single qubit by the deformation, giving a selective spontaneous
radiation spectrum. Further, unlike a microscopic lattice with large-$N$
limit and nearly zero relative spacing, the polariton in the quasi-lattice
has uneven decay rate over the relative spacing. We show that this
unevenness coincides with the cooperative emission effect expected
from the superradiance model, where alternative excitations in the
qubits of the lattice result in maximum decay.
\end{abstract}


\maketitle

\section{Introduction}

\subsection{Quasi-lattices of qubits}

Superconducting qubits are a class of two-level systems based on the
superconducting Josephson junctions~\cite{clarke08}. When interacting
with a microwave field stalled in a stripline resonator, they act
like artificial atoms in optical cavities with coherent exchange of
photon energy modeled on the Jaynes-Cummings Hamiltonian~\cite{wallraff04,majer07}.
This combination gives rise to circuit quantum electrodynamics (QED)~\cite{blais04},
an emulation of cavity QED. Circuit QED systems with superconducting
qubits can therefore emulates, in many aspects, quantum optical effects
similar to those originally discovered on real atoms or atomic media~\cite{jqyou11}
and can be regarded as a type of quantum simulators~\cite{buluta09}.

So far, studies of circuit QED only concerns with circuits of a few
superconducting qubits and each qubit is considered acting separately
with the stripline resonator. To understand their collective behavior,
a many-qubit theory is wanted. However, traditional many-atom theories
such as the Frenkel model for excitons cannot not be applied because
circuit QED systems contain only a finite number of artificial atoms,
i.e., large-$N$ limit is not taken. Neither do models for finite
$N$, such as the Tavis-Cummings (TC) model~\cite{tavis68} apply
because the mesoscopic sizes of the qubits and of the spacings between
the qubits in a circuit make their dipole-field interaction to the
stripline resonator inhomogeneous. To remedy the inapplicability,
we have proposed a projection-deformation model that generalizes the
TC-model to work on inhomogeneous coupling scenarios~\cite{ian12,ian13}.

The basic idea is that while real atoms form a lattice, the superconducting
qubits form a quasi-lattice, whose ``quasi-ness'' is measured by the
degree of deformation it departs from a normal lattice. This deformation
is related to a deformed SU(2) algebra~\cite{sun89} obeyed by the
collective spin operators of the quasi-lattice and is quantified by
a c-number deformation factor. This factor is parametrized by the
ratio of the uniform lattice spacing $L_{\mathrm{q}}$ to the wavelength
$\lambda_{\mathrm{p}}$ of the interacting photon. We call this ratio
the relative spacing $\ell$ of the qubits. The analytical excitation
spectrum of the quasi-lattice can then be computed by diagonalizing
the deformation-dependent interaction Hamiltonian.

\subsection{Polariton and the radiation problem}

The excited states of the quasi-lattice ought to spontaneously radiate
microwave photons into the circuit waveguide. But in a circuit QED
system, the excitation mode of the quasi-lattice is coherently coupled
to a stripline resonator mode, where the mixed mode of the two forms
a cavity polariton. Therefore, unlike the conventional radiation problems
treated in atomic physics, such as the Dicke radiation~\cite{dicke54,ernst68},
the radiation in the mesoscopic circuit cavity is associated with
the dynamics of the polariton, i.e. a \emph{dressed} quasi-lattice
excitation mode, instead of a bare lattice excitation mode.

It is known that, in a dielectric, the polariton mode is generated
by the recombination of the collective atomic excitations with the
radiated photons~\cite{knoester89}. The polariton mode we treat
here, on the other hand, is generated by the dressing through the
cavity photon. The definition is similar to that given in semiconductor
cavity QED systems~\cite{keeling07,savona94}. The radiated photon
modes are designated in a separate Hilbert space from that of the
cavity photon mode. Therefore, even though the radiation processes
are both polariton-mediated, the case for a quasi-lattice of qubits
is vastly different from the case of a dielectric.

Besides, the energies and the eigenstates of the quasi-lattice excitation
modes are modified by the deformation described above. The composition
of the Fock number states of the polariton depends thus not only on
the eigenenergy of the qubits, but also on the deformation of the
quasi-lattice. It was shown that the structure of the underlying medium
has a large influence on the pattern and form of the radiation. For
example, the spontaneous emission is found directionally dependent
on the incident photon~\cite{scully06} in an extended medium; uneven
decay rates is found in a spherical symmetric medium by including
virtual photon processes~\cite{svidz10}; and nonlocal effect arises
for single-photon cooperative emission~\cite{svidz12}. The purpose
of this paper is to investigate how the structural change introduced
by the deformation of the quasi-lattice affects the radiation spectrum
and the polariton decay.

We find that the cavity polariton has an acutely \emph{selective}
distribution of its radiated microwave photon based on the deformation.
This is shown by the varying amplitude of the interaction coefficient
between the polariton and the continuum of photon modes in the momentum
space. This amplitude depends on the relative spacing and has a quasi-periodicity
that matches this relative spacing with the wavelength of the cavity
photon, demonstrating the selectiveness of the quasi-lattice about
its radiation. Specifically, at the exact periodic positions where
the radiated photon resonates with the cavity, the magnitude of the
interaction will obtain its maximum value.

Moreover, we find that the polariton decay is also deformation-dependent
on the relative spacing. In fact, as predicted by Dicke, the decay
rate of an $N$-atom lattice would increase to $N^{2}$ when the spin
moment of the lattice is at the maximum cooperation number of $N/2$,
giving rise to superradiance~\cite{rehler71,macgill76,ressayre77}
and superfluorescence~\cite{bonifacio75,glauber78,polder79}. For
the quasi-lattice of qubits, it is found that the maximum decay rate
is obtained when the relative spacing is set to one half, where only
every other qubit couples to the cavity photon. This alternate pattern
of coupling excites half of the qubits while leaving the other half
unaffected, giving an effective spin moment of $N/2$ to the quasi-lattice
and having the decay rate match with the Dicke model of cooperated
radiation.

The article is organized as follows. The formation of cavity polariton
in a quasi-lattice is given in Sec.~\ref{sec:Polaritons}, where
the transition matrices for the quasi-lattice as deformed SU(2) spin
is derived in the polariton basis. By writing the qubit operators
terms of these matrices using a discrete Fourier transform, we derive
the expression of the quasi-periodic interaction coefficient for radiation
in Sec.~\ref{sec:EoM}A. As an example, the simplest non-trivial
case with $N=4$ is plotted especially to illustrate the uneven distribution
of radiation of the mesoscopic system. With the derived interaction
coefficient, the equation of motion for the low-energy polariton states
are derived in Sec.~\ref{sec:EoM}B. The decay rate of the polariton
is subsequently computed under the Markov and the Wigner-Weisskopf
approximations in Sec.~\ref{sec:Soln}. The conclusion and relevant
discussions are given in Sec.~\ref{sec:Conclusion}.

\section{Polaritons\label{sec:Polaritons}}

\subsection{System state space}

\begin{figure}
\includegraphics[clip,width=8.5cm]{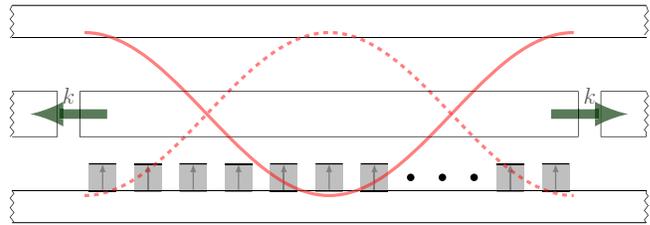}

\protect\caption{(Color online) Schematic of a quasi-lattice chain of qubits coupled
to a cavity field. The arrow $k$ indicates the direction of photons
radiated by the polaritons formed from the excitation of the quasi-lattice
and the cavity field.\label{fig:model}}
\end{figure}

Consider the circuit QED system illustrated in Fig.~\ref{fig:model}
with $N$ superconducting qubits (indicated by gray squares), where
two neighboring ones are spaced at a uniform distance $L_{\mathrm{q}}$.
Each qubit can be modeled as a two-level system $\sigma_{j,z}$ in
the diagonalized basis of the Josephson and charge energies of the
junctions that it contains. Depending on the type of the qubit, the
diagonalized eigenenergy is tunable through magnetic flux, gate charge,
phase, etc. and we consider the $N$ qubits are tuned uniform with
level spacing $\omega_{\mathrm{q}}$ in our study here.

The middle rectangular strip indicates the coplanar waveguide or stripline
resonator, which is equivalent to a cavity and contains multiple modes
of a standing microwave field. However, when the qubits are all tuned
resonant with the fundamental mode, the cavity field can be effectively
regarded as a single-mode field~\cite{fink09,macha13}. We describe
this fundamental mode (the red curve in the figure) by the annihilation
operator $\bar{a}$ and denote its frequency and wavelength by $\omega_{\mathrm{C}}$
and $\lambda_{\mathrm{C}}$, respectively. Note that even though the
dimensions of a qubit are negligible compared to the wavelength $\lambda_{\mathrm{C}}$,
the spacings between the qubits are non-negligible and the coupling
of each qubit to the cavity field depends on the relative spacing
$\ell=2L_{\mathrm{q}}/\lambda_{\mathrm{C}}$ described above through
a sinusoidal factor $\cos(j\pi\ell)$. The variable $j$ consequently
can be regarded as a relative coordinate for the qubits along the
one-dimensional chain of qubits. This chain of qubits can be regarded
as a quasi-lattice, which resembles an atomic lattice but has an inhomogeneous
coupling due to its mesoscopic nature.

This quasi-lattice of qubits is also environmentally coupled to a
reservoir, represented by a continuum of quantum oscillators $\{a_{k}\}$
with frequency spectrum $\omega_{k}$, which gives rise to spontaneous
radiation in and out of the waveguide. These radiated photons carry
momentum $k$ and are illustrated as the green arrows in Fig.~\ref{fig:model}.
Their propagations in the waveguide are indicated by the wave functions
$e^{ikr_{j}}$, where $r_{j}$ is the coordinate of the associated
qubit. Note that, unlike the treatments for radiation in atomic media
where confinements in pencil-shape geometries are usually assumed~\cite{ressayre77,bonifacio75},
the waveguide in superconducting circuits are strictly one-dimensional
and thus $k$ is regarded as a wave number, not a wave vector.

The total system Hamiltonian is therefore divided into three parts
and can be written as follows (assuming $\hbar=c=1$)
\begin{eqnarray}
H & = & H_{\mathrm{sys}}+V_{\mathrm{cav}}+V_{\mathrm{rad}},\\
H_{\mathrm{sys}} & = & \omega_{\mathrm{q}}\sum_{j=0}^{N-1}\sigma_{j,z}+\omega_{\mathrm{C}}\bar{a}^{\dagger}\bar{a},\label{eq:sys_Ham}\\
V_{\mathrm{cav}} & = & \eta\sum_{j=0}^{N-1}\cos(j\pi\ell)\left[\sigma_{j,+}\bar{a}+\sigma_{j,-}\bar{a}^{\dagger}\right],\label{eq:int_Ham}\\
V_{\mathrm{rad}} & = & \sum_{j=0}^{N-1}\sum_{k}g_{k}\left(a_{k}\sigma_{j,+}e^{ikr_{j}}+a_{k}^{\dagger}\sigma_{j,-}e^{-ikr_{j}}\right).\label{eq:rad_Ham}
\end{eqnarray}
Note that the forms taken by Eqs.~(\ref{eq:sys_Ham})-(\ref{eq:int_Ham})
assumes a strong coupling operation regime for the qubits, where each
qubit has a maximal coupling strengh $\eta$ much greater than the
linewidth of the circuit cavity, to ensure coherent exchange of photons
with the circuit cavity while the non-rotating wave terms for the
virtual photons can be ignored~\cite{majer07}. Further, based on
the current experiment setups~\cite{macha13}, the magnitude of $\eta$
is much smaller than $\omega_{\mathrm{q}}$ such that the quasi-lattice
system on the other hand does not enter into the ultra-strong coupling
(USC) regime. When $\eta$ is comparable to $\omega_{\mathrm{q}}$,
USC operation will dominate and squeezing terms of $a$ and $a^{\dagger}$
have to be taken into considerations~\cite{deliberato14}. These
terms entail complex implications to Dicke phase transitions on circuit
QED systems~\cite{nataf10,viehmann11}.

The relevant Hilbert space is tripartite: \emph{
\begin{equation}
\mathcal{H}=\mathcal{Q}^{\otimes N}\times\mathcal{U}\times\prod^{\otimes k}\mathcal{V}_{k},\label{eq:Hilbert_space}
\end{equation}
}where each $\mathcal{Q}=\mathrm{span}\{\left|\uparrow\right\rangle ,\left|\downarrow\right\rangle \}$
is the internal energy eigenspace for a qubit, $\mathcal{U}=\mathrm{span}\{\left|0\right\rangle ,\left|1\right\rangle ,\left|2\right\rangle ,\dots\}$
is the Fock eigenspace for the cavity photon, and each $\mathcal{V}_{k}=\mathrm{span}\{\left|0\right\rangle _{k},\left|1\right\rangle _{k},\left|2\right\rangle _{k},\dots\}$
is the Fock eigenspace for spontaneous emitted photon of wave number
$k$. A specific system state vector, for example, in this tripartite
Hilbert space can be written as
\begin{equation}
\left|\downarrow\uparrow\downarrow\downarrow\cdots\uparrow\right\rangle \otimes\left|n\right\rangle \otimes\left|\{k_{j}\}\right\rangle ,\label{eq:prod_state}
\end{equation}
where the first subvector denotes the configuration of the quasi-lattice
of qubits, the second that of cavity photon number, and the third
that of the photon momentum radiated by each qubit in the quasi-lattice.

Following the idea of either Dicke or Tavis-Cummings, the quasi-lattice
can be equally expressed in the angular momentum space $\left|r,m\right\rangle $
where $r=N/2=\frac{1}{2}(n_{\uparrow}+n_{\downarrow})$ is the total
quantized spin and $m$ the magnetic moment. $m$ is also the difference
between the number of spin-up qubits and the number of spin-down qubits
in the quasi-lattice, i.e. $m=\frac{1}{2}(n_{\uparrow}-n_{\downarrow})\in\{-r,-r+1,\dots,r\}$.
Hence, a quasi-lattice state expressed in $\left|r,m\right\rangle $
space has the following correspondence to the qubit spin states
\begin{equation}
\left|r,m\right\rangle =\sqrt{\frac{(r+m)!(r-m)!}{2r!}}\sum_{p}\Pi_{p}\left|\underbrace{\uparrow\uparrow\cdots\uparrow}_{r+m}\underbrace{\downarrow\downarrow\cdots\downarrow}_{r-m}\right\rangle ,\label{eq:ang_momt_state}
\end{equation}
where $\Pi_{p}$ is a permutation operation on the ordered spin state
$\left|\uparrow\uparrow\cdots\uparrow\downarrow\downarrow\cdots\downarrow\right\rangle $
and the summation is over all permutations of the same $m$. In other
words, the number of permutations is the degeneracy of the state $\left|r,m\right\rangle $,
which is just the reciprocal of the constant in front of the summation
in Eq.~(\ref{eq:ang_momt_state}).

Corresponding to this angular momentum space representation of the
state of the quasi-lattice, we introduce a set of total angular momentum
operators
\begin{eqnarray}
S_{z} & = & \sum_{j=0}^{N-1}\sigma_{j,z},\\
S_{+} & = & \sum_{j=0}^{N-1}\cos(j\pi\ell)\sigma_{j,+},\label{eq:S_+}
\end{eqnarray}
and $S_{-}=(S_{+})^{\dagger}$ to replace the Pauli operators for
the individual qubits. These operators obey the structure of a deformed
SU(2) algebra~\cite{sun03}, i.e. we have the commutator become 
\begin{equation}
[S_{+},S_{-}]=2\Sigma_{z}
\end{equation}
where $\Sigma_{z}$ is an operator function of $S_{z}$ and $\sigma_{j,z}$
with coefficients depending on $N$ and the relative spacing $\ell$~\cite{ian12}.
When $N\to\infty$ and $\ell\to0$, the coefficients in front of $\sigma_{j,z}$
vanish and $\Sigma_{z}=S_{z}$, for which the normal SU(2) algebra
is reinstalled. With the introduction of these operators, the expressions
for Hamiltonians (\ref{eq:sys_Ham})-(\ref{eq:int_Ham}) can be simplified
to
\begin{eqnarray}
H_{\mathrm{sys}} & = & \omega_{\mathrm{q}}S_{z}+\omega_{\mathrm{C}}\bar{a}^{\dagger}\bar{a}\label{eq:sys_Ham_tfm'd}\\
V_{\mathrm{cav}} & = & \eta(S_{+}\bar{a}+S_{-}\bar{a}^{\dagger})\label{eq:int_Ham_tfm'd}
\end{eqnarray}

\subsection{Diagonalizing for polariton}

The polariton state is the eigenstate that diagonlizes $H_{\mathrm{sys}}+V_{\mathrm{cav}}$
by transforming the first two product spaces in Eq.~(\ref{eq:Hilbert_space}).
It arises as the dressed state the quasi-lattice excitation by the
cavity field.

Since the coupling of the quasi-lattice to the cavity is inhomogeneous,
the polariton state contains implicitly a dependence on the relative
spacing $\ell$. Following the projection-deformation (PD) method
we have introduced~\cite{ian12,ian13}, it can be written as the
eigenstate
\begin{equation}
\left|u,r\right\rangle =\sum_{n}c_{n}^{(u)}\left|r,u-n;n\right\rangle ,\label{eq:expansion}
\end{equation}
where $u$ denotes the total excitation number. This number $u$ is
shared between $n$ for the photon energy part and $m=u-n$ for the
quasi-lattice excitation part. It can take either integer or half-integer
values since $m$ can be half-integer for odd $N$ of qubits. The
implicit dependence on $\ell$ is reflected in the expansion coefficients
\begin{equation}
c_{n}^{(u)}=\mathscr{P}_{n}^{(u)}\sum_{q=0}^{n/2}\mathfrak{f}^{q-\frac{n}{2}}\mathscr{C}_{q}^{(u)},\label{eq:coeff}
\end{equation}
through the deformation factor
\begin{equation}
\mathfrak{f}=\frac{1}{2}+\frac{1}{4N}\left(1+\frac{\sin(2N-1)\pi\ell}{\sin\pi\ell}\right).\label{eq:defm_factor}
\end{equation}

The expression for coefficients in Eq.~(\ref{eq:coeff}) is found
by solving a recursive relation. Recursively expanding the relation,
each iteration gives a term that has the same factor 
\begin{equation}
\mathscr{P}_{n}^{(u)}=\frac{\prod_{j=0}^{n-1}\left[\varepsilon-j\Delta\omega\right]}{\sqrt{n!(r+u)^{\underline{n}}(r-u+1)^{\bar{n}}}},
\end{equation}
where $\varepsilon$ is eigenvalue for the interaction $V_{\mathrm{cav}}$
in Eq.~(\ref{eq:int_Ham}) and $\Delta\omega=\omega_{\mathrm{C}}-\omega_{\mathrm{q}}$
is the qubit-cavity detuning. We have used the Pochhammer symbols
$x^{\underline{n}}=x(x-1)\cdots(x-n+1)$ and $x^{\bar{n}}=x(x+1)\cdots(x+n-1)$
to simplify the notation. The factor $\mathscr{C}_{q}^{(u)}$ with
a fixed $q$ can be regarded as the contribution to a $u$-number
excitation mode with $q$ parts of excitation from in the quasi-lattice
alone. Written explicitly, it reads 
\begin{multline}
\mathscr{C}_{q}^{(u)}=\underset{\langle j_{1}\dots j_{k}\dots j_{n/2}\rangle}{\sum\cdots\sum}\prod_{k=1}^{q}(-\eta^{2})(j_{k}+1)\times\\
\frac{r+u-j_{k}}{v-j_{k}\Delta\omega}\cdot\frac{r-u+j_{k}+1}{v-(j_{k}+1)\Delta\omega},
\end{multline}
where ,under the multi-dimensional summation, $\langle j_{1}\dots j_{k}\dots j_{n/2}\rangle$
represents the descending index set $\{\forall k<l:0\leq j_{l}\leq j_{k}-2\}$.
For example, for $\frac{n}{2}=2$, the summation is two dimensional,
with the first index $j_{1}\in\{0,1,2\}$ and the second index $j_{2}\in\{0\}$.
Appendices~B and C of Ref.~\cite{ian12} gives the detailed derivation.

\subsection{Excitation operators in polariton basis}

The vector $\left|r,m;n\right\rangle $ represents the state of the
quasi-lattice and cavity system by denoting the lattice excitation
and the photon state, separately. Whereas the vector $\left|u,r\right\rangle $
represents the same combination by denoting the polariton state. Consequently,
the part of the total Hamiltonian not relating to spontaneous radiation,
i.e. Eqs.~(\ref{eq:sys_Ham_tfm'd})-(\ref{eq:int_Ham_tfm'd}), can
be written in the polariton basis
\begin{equation}
H_{\mathrm{sys}}+V{}_{\mathrm{cav}}=\sum_{u}\Omega_{u}\left|u,r\right\rangle \left\langle u,r\right|\label{eq:polariton}
\end{equation}
where the eigenfrequency $\Omega_{u}$ is determined \emph{a posteriori}
by a recursive relation~\cite{ian13}. The ladder operators $S_{+}$
and $S_{-}$ in $V_{\mathrm{cav}}$, originally indicating the collective
excitation from individual qubits, should now be written as the off-diagonal
elements of the transition matrix in the transformed basis as well.

To find the expression of the matrix elements, we expand the bra's
and ket's of the polariton state vector into the two-partite form
in Eq.~(\ref{eq:expansion}). We can observe that, even though there
is exchange of energy between the quasi-lattice and the cavity field,
the total number $u$ of excitations is preserved over the exchange
process when we disregard the energy gain and loss due to spontaneous
radiation, as reflected in the interaction of Eq.~(\ref{eq:rad_Ham}).
As a result, the diagonal elements of the transition matrices for
operators $S_{+}$ and $S_{-}$ are zero in the polariton basis, as
we have verified in App.~\ref{sec:Deriving-excitation-operator}.

For the non-diagonal elements, we first observe that the non-uniformity
of the quasi-lattice has the effect of reducing transition amplitudes
as photons are more difficult to be either absorbed or emitted with
$\mathfrak{f}\leq1$:
\begin{multline}
S_{+}\left|r,u-n;n\right\rangle \\
=\sqrt{\mathfrak{f}(r-u+n)(r+u-n+1)}\left|r,u-n+1;n\right\rangle ,\label{eq:S+_op}
\end{multline}
\begin{multline}
S_{-}\left|r,u-n;n\right\rangle \\
=\sqrt{\mathfrak{f}(r+u-n)(r-u+n+1)}\left|r,u-n-1;n\right\rangle .\label{eq:S-_op}
\end{multline}
Applying the two operation rules above, the matrix elements in the
polariton basis vanish except for the first off-diagonal line because
of the conservation of energy in the total excitation number $u$.
This gives the raising operator as a lower off-diagonal matrix
\begin{equation}
[S_{+}]_{u,u-1}=\sum_{n}c_{n}^{(u)}c_{n}^{(u-1)}\sqrt{\mathfrak{f}(r+u-n)(r-u+n+1)},\label{eq:matrix_S+}
\end{equation}
and the lowering operator as an upper off-diagonal matrix
\begin{equation}
[S_{-}]_{u,u+1}=\sum_{n}c_{n}^{(u)}c_{n}^{(u+1)}\sqrt{\mathfrak{f}(r-u+n)(r+u-n+1)}.\label{eq:matrix_S-}
\end{equation}

\begin{figure}
\includegraphics[bb=0bp 0bp 380bp 218bp,width=7.2cm]{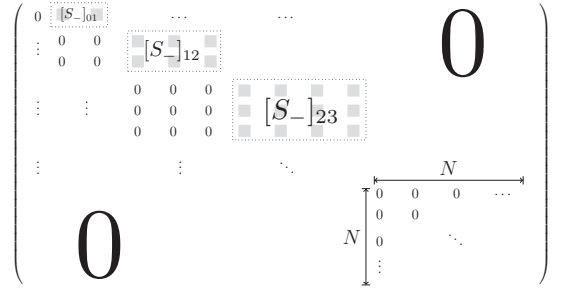}

\protect\caption{Matrix representation of the ladder operator $S_{-}$ in the photon
basis, where the non-zero elements are clustered in blocks off the
block-diagonal line.\label{fig:matrix} }
\end{figure}

Since each polariton consists of a series of combinations of photons
and quasi-lattice excitations that sum up to the same total excitation
number, the $(u,u-1)$-th non-zero element of $S_{+}$ in the polariton
basis can be expanded as a $u+1$ by $u$ block submatrix in the photon
basis of $n$. This submatrix resides under the $(u,u)$-th block
matrix along the diagonal, making the transition matrix $S_{+}$ lower
block off-diagonal. Corresponding, $S_{-}$ in the photon basis is
upper block off-diagonal, where the $(u,u+1)$-th element expands
to a $u+1$ by $u+2$ submatrix, as shown in Fig.~\ref{fig:matrix}.

\section{Equations of motion\label{sec:EoM}}

\subsection{Discrete Fourier transform and coupling coefficients}

Equipped with Eqs.~(\ref{eq:matrix_S+})-(\ref{eq:matrix_S-}), we
are ready to deal with the Hamiltonian (\ref{eq:rad_Ham}) responsible
for radiation. First, we need to consider the qubit operators $\sigma_{j,+}$
and $\sigma_{j,-}$ in the polariton space. This can be done by regarding
the operator $S_{+}$ defined in Eq.~(\ref{eq:S_+}) as a discrete
cosine transform of $\{\sigma_{j,+}\}$. Each $\sigma_{j,+}$ can
then be written as the inverse transform
\begin{equation}
\sigma_{j,+}=\frac{1}{N}\sum_{Nl=0}^{N-1}\cos(j\pi l)S_{l,+},\label{eq:sgm_+_Fourier}
\end{equation}
where we have used $Nl\in\{0,\cdots,N-1\}$ to denote the index $l\in\{0,\frac{1}{N},\dots,\frac{N-1}{N}\}$.
$j$ and $l$ hence become a pair of conjugate variables for the discrete
Fourier transforms such that (\ref{eq:S_+}) and (\ref{eq:sgm_+_Fourier})
satisfy the orthonormality and unitarity conditions imposed by Parseval's
theorem.

Interpreted physically, the forward transform regards that the individual
qubit excitations over all $j$ positions constitute the collective
excitation, where those with $j$ at the antinodes of the cavity field
contribute most to the amplitude of the collective excitation. Whereas,
the inverse transform implies that the collective excitations over
a set of particular $l$ constitute an individual excitation at $j$,
where the more $l$ matches with $j$, the more it will contribute
to the amplitude of the individual excitation.

We should emphasize that even though $l$ designates length, its meaning
is distinct from $\ell$. While $\ell=2L_{\mathrm{q}}/\lambda_{\mathrm{p}}$
is a fixed value determined by the physical circuit layout, $l$ is
only an indexing or transform variable that takes value from a discrete
set of numbers.

Substituting the inverse transform of Eq.~(\ref{eq:sgm_+_Fourier})
into Eq.~(\ref{eq:rad_Ham}), we have the Hamiltonian in the polariton
basis

\begin{multline}
H=\sum_{u}\Omega_{u}\left|u,r\right\rangle \left\langle u,r\right|+\sum_{k}\omega_{k}a_{k}^{\dagger}a_{k}+\\
\sum_{j,Nl=0}^{N-1}\sum_{k}\frac{g_{k}}{N}\left[a_{k}S_{l,+}\cos(j\pi l)e^{ikr_{j}}+\mathrm{h.c.}\right],
\end{multline}
for which the system now essentially consists of two parts: the polaritons
and the radiated photons from the polaritons. Since the collective
excitation operator $S_{l,+}$ has no dependence on the relative coordinate
$j$, the functions involving $j$ in the second line of the equation
can be summed. Writing the coordinates $r_{j}=jL_{\mathrm{q}}=j\ell\pi/k_{0}$
where $k_{0}=2\pi/\lambda_{\mathrm{C}}$ is the momentum of the cavity
photon, we find the radiation part of the Hamiltonian become
\begin{equation}
V_{\mathrm{rad}}=\sum_{Nl=0}^{N-1}\sum_{k}\frac{g_{k}}{N}\left[\chi_{l}(k)a_{k}S_{l,+}+\mathrm{h.c.}\right]\label{eq:tfm_int_Ham}
\end{equation}
where the coefficient
\begin{multline}
\chi_{l}(k)=\biggl[1+e^{i\ell\pi(N+1)k/k_{0}}\cos l(N-1)\pi\\
-e^{i\ell N\pi k/k_{0}}\cos Nl\pi-e^{i\ell\pi k/k_{0}}\cos l\pi\biggr]\Bigl/\\
\biggl[1+e^{2i\ell\pi k/k_{0}}-2e^{i\ell\pi k/k_{0}}\cos l\pi\biggr]\label{eq:coeff_chi}
\end{multline}
is a function of the momentum $k$ of the radiation photon.

Comparing the interaction term of Eq.~(\ref{eq:tfm_int_Ham}) with
that of Eq.~(\ref{eq:rad_Ham}), we observe the original interaction
between the individual qubits and the radiation is quantified by the
dipole-field coupling amplitude $g_{k}$ only, which depends on the
dipole moment of the qubits and the volume of the cavity. The coefficient
$\chi_{l}(k)$ shows that an extra gain factor is introduced because
the photons here are radiated from the polaritons instead of the individual
qubits. Since the polaritons arises from the resonance between the
quasi-lattice and the cavity, this factor $\chi_{l}(k)$ is determined
by the geometric structure of the underlying quasi-lattice: the total
number of qubits $N$ and the relative spacing $\ell$.

\begin{figure}
\includegraphics[bb=107bp 263bp 492bp 555bp,clip,width=8cm]{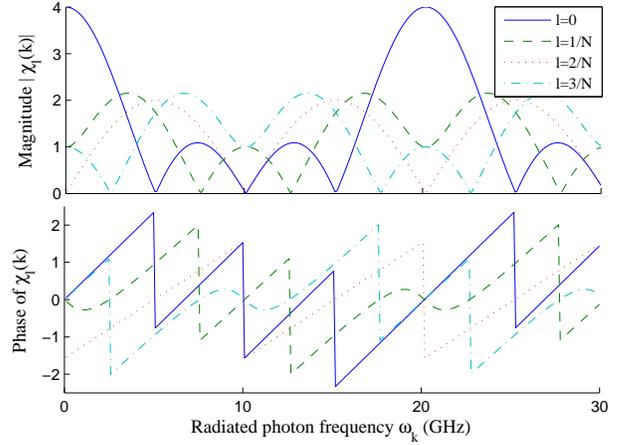}

\protect\caption{(Color online) (a) The magnitude and (b) the phase of the periodic
coupling coefficient $\chi_{l}(k)$, for photons radiated by polaritons
in the circuit QED system over a frequency range $0<\omega_{k}=ck<30$GHz.
A quasi-lattice of $N=4$ qubits with relative spacing $\ell=2/3$
is assumed, for which $l$ can take four values $0$, $1/4$, $1/2$,
and $3/4$. \label{fig:coupling_coeff}}
\end{figure}

How $N$ and $\ell$ determines the radiation character of the mesoscopic
system can be illustrated from the simplest non-trivial example with
an $N=4$ quasi-lattice, which is experimentally realizable in circuit
QED~\cite{fink09}. We take $\ell$ to be $2/3$ for a typical inhomogeneous
coupling. Since $\chi_{l}$ is a complex number, its magnitude and
phase are plotted separately against the radiated photon frequency
in Fig.~\ref{fig:coupling_coeff}. In the Figure, we have assumed
$k_{0}$ adopt the value based on the first harmonic given in the
multi-qubit circuit by Fink \emph{et al}~\cite{fink09}, where $\omega_{\mathrm{C}}=ck_{0}=6.729$GHz.

We observe that the periodicity of the many exponentials in Eq.~(\ref{eq:coeff_chi})
makes $\chi_{l}(k)$ quasi-periodic. The quasi-period is 
\begin{equation}
K=\frac{2k_{0}}{\ell}=\frac{\lambda_{\mathrm{C}}}{L_{\mathrm{q}}}k_{0}.\label{eq:period}
\end{equation}
For the $\ell=2/3$ case plotted in the figure, the period in terms
of frequency is $\omega_{K}=20.2$GHz. This quasi-periodicity matches
the zeros of $|\chi_{l}|$ with the discontinuities of the phase of
$\chi_{l}$ and the local maxima of $|\chi_{l}|$ with the zero phase
of $\chi_{l}$. In fact, since $\chi_{l}(k)$ is an entire function
of the real variable $k$, the real and the imaginary parts of $\chi_{l}(k)$
obey the Kramers-Kronig relations if we extend $k$ to the complex
plane. This implies an anomalous absorption and dispersion relation
of the radiation spectrum of the polariton due to the inhomogeneous
coupling~\cite{stephen64}.

The radiation process is therefore either enhanced or surpressed,
depending on whether how well the radiation photon of frequency $\omega_{k}$
matches with the resonance between the cavity of frequency $\omega_{\mathrm{C}}$
and the quasi-lattice of relative spacing $\ell$. The fact that certain
radiation range can be enhanced is because the photons emitted from
these ranges are reabsorbed by the quasi-lattice before reemitting
into the waveguide. From Eqs.~(\ref{eq:coeff_chi})-(\ref{eq:period}),
the number of extrema in $\chi_{l}$, indicating the exact matching
and the exact mismatching, is determined by the number of qubits $N$
while the distance in $k$ between two such extrema is determined
by the relative spacing $\ell$. 

In other words, we expect the polariton radiation on a quasi-lattice
of superconducting qubits provides a selective spectrum of radiation
according not only to the eigenenergy of the qubits, but also to their
geometric layout in the circuit. The quasi-periodic character is unique
to the mesoscopic nature of the circuit QED system. For in an atomic
lattice, the equivalent inter-atom spacing $L_{\mathrm{q}}$ approaches
zero. The corresponding quasi-period $K$ defined by Eq.~(\ref{eq:period})
will approach infinity and the periodicity vanishes.

\subsection{Equations of motion}

To study how exactly the varying interaction coefficient affects the
radiation spectrum, we derive the equation of motion of the polariton
here and compute its decay rate in the following section.

Introducing the eigenfrequency $\omega_{k}=ck$ for the radiation
photon, the Hamiltonian in the interaction picture between the polariton
and the radiation photon reads (see App.~\ref{sec:comm_relations}
for discussion)
\begin{equation}
H'_{\mathrm{int}}(t)=\sum_{Nl=0}^{N-1}\sum_{k}\frac{g_{k}}{N}\left[\chi_{l}(k)a_{k}S_{l,+}e^{-i(\omega_{\mathrm{q}}-\omega_{k})t}+\mathrm{h.c.}\right].\label{eq:Ham_int-pic}
\end{equation}
To study the low-energy dynamics of multi-atom systems with weak coupling,
it is customary to look at the Schroedinger equations of the lowest
excited product states: states with either one excited atom across
the lattice and zero photon or no excited atom and one photon~\cite{scully06,svidz10,svidz12}.
In the polariton basis, these two states translate into the $\mathbf{0}$-excitation
state and the $\mathbf{1}$-excitation state of the dressed quasi-lattice
\begin{equation}
\left|\psi(t)\right\rangle =\alpha(t)\left|\mathbf{1},r\right\rangle \otimes\left|0\right\rangle +\sum_{k}\beta_{k}(t)\left|\mathbf{0},r\right\rangle \otimes\left|1_{k}\right\rangle ,\label{eq:state_vec}
\end{equation}
which couple, respectively, to the zero and the one radiated photon
state. The radiated photons are indexed by their momentum $k$.

Applying Schrödinger equation with the Hamiltonian~(\ref{eq:Ham_int-pic})
to this state, we find a pair of coupled equations of motion for the
two coefficients of the state vector
\begin{eqnarray}
\frac{\mathrm{d}\alpha}{\mathrm{d}t} & = & -i\sum_{Nl=0}^{N-1}\sum_{k}\frac{g_{k}}{N}\beta_{k}\chi_{l}(k)[S_{l,+}]e^{-i(\omega_{\mathrm{q}}-\omega_{k})t}\label{eq:EoM_alpha}\\
\frac{\mathrm{d}\beta_{k}}{\mathrm{d}t} & = & -i\sum_{Nl=0}^{N-1}\frac{g_{k}}{N}\alpha\chi_{l}^{\ast}(k)[S_{l,-}]e^{i(\omega_{\mathrm{q}}-\omega_{k})t}\label{eq:EoM_beta}
\end{eqnarray}
where $[S_{+}]$ and $[S_{-}]$ are the matrix elements given by Eq.~(\ref{eq:matrix_S+})-(\ref{eq:matrix_S-})
for the transitions between $\mathbf{0}$-excitation state and $\mathbf{1}$-excitation
state.

We note here that the coupled equations of motion are not dissimilar
to those originally given for the radiation of atomic lattices. The
fact that it is the polaritons undergoing the radiation is reflected
by two changes to the original equations: (i) the amplitude of radiation
is controlled by the extra factor $\chi_{l}(k)$ and (ii) the radiated
photon is not contributed by a single excitation, but by a group of
them with non-zero transition probabilities distributed in $[S_{l,\pm}]$.
The weight of each contribution is determined by the deformation factor
$\mathfrak{f}$ and thus by the relative spacing $\ell$ of the quasi-lattice.

To solve the coupled equations of polariton dynamics, consider that
the dipole-field exchange of excitation of each qubit is a much slower
process than the photon oscillation in the cavity. That means mathematically
the variation of the function $\alpha(t)$ is adiabatic compared to
the propagating functions $e^{i\omega_{k}t}$ and $e^{i\omega_{\mathrm{q}}t}$
of the photons. In other words, the change of $\alpha(t)$ can be
computed from the average of the temporal exponentials. We hence combine
Eqs.~(\ref{eq:EoM_alpha})-(\ref{eq:EoM_beta}) and apply Markov
approximation to arrive at the equation
\begin{equation}
\dot{\alpha}=-\sum_{k}g_{k}^{2}|\mathfrak{s}(k)|^{2}\alpha(t)\int_{0}^{t}\mathrm{d}\tau\, e^{-i(\omega_{\mathrm{q}}-\omega_{k})(t-\tau)},\label{eq:EoM_Markov}
\end{equation}
where we have used the notation 
\begin{equation}
\mathfrak{s}(k)=\frac{1}{N}\sum_{Nl=0}^{N-1}\chi_{l}(k)[S_{l,+}].\label{eq:frak_s}
\end{equation}
The application of Markov approximation also accords with Wigner-Weisskopf's
original treatment of atomic decay, by assuming the decaying function
be exponential. That is, the exponetial function $\alpha(\tau)\approx e^{-\gamma\tau/2}$
would be a much slower process than the oscillating function $e^{-i\omega\tau}$.
We follow this line of thought and derive the decay of polariton in
the next section.

\section{Wigner-Weisskopf approximation and decay rate\label{sec:Soln}}

\subsection{Wigner-Weisskopf approximation}

The approximation of Wigner-Weisskopf assumes a continuous radiation
spectrum, so the summation over momentum $k$ can be extended to an
integral and Eq.~(\ref{eq:EoM_Markov}) reads
\begin{equation}
\frac{\mathrm{d}}{\mathrm{d}t}\alpha(t)=-\frac{\lambda_{\mathrm{C}}}{2\pi}\int\mathrm{d}k\, g_{k}^{2}|\mathfrak{s}(k)|^{2}\alpha(t)\frac{1-e^{-i(\omega_{\mathrm{q}}-\omega_{k})t}}{i(\omega_{\mathrm{q}}-\omega_{k})}.\label{eq:EoM_WW}
\end{equation}

At steady state ($t\to\infty$), the last factor with the exponential
can be replaced by a principal value and a delta function
\begin{equation}
\frac{1-e^{-i(\omega_{\mathrm{q}}-\omega_{k})t}}{i(\omega_{\mathrm{q}}-\omega_{k})}=-i\frac{\mathscr{P}}{\omega_{\mathrm{q}}-\omega_{k}}+\pi\delta(\omega_{\mathrm{q}}-\omega_{k}).\label{eq:zeta_defn}
\end{equation}
Replacing this factor into Eq.~(\ref{eq:EoM_WW}), we get
\begin{equation}
\frac{\mathrm{d}}{\mathrm{d}t}\alpha(t)=\alpha(t)\frac{\lambda_{\mathrm{C}}}{2\pi c}\left\{ i\mathscr{P}\negthickspace\int\mathrm{d}k\frac{g_{k}^{2}|\mathfrak{s}(k)|^{2}}{k_{\mathrm{q}}-k}-\pi g_{k_{\mathrm{q}}}^{2}|\mathfrak{s}(k_{\mathrm{q}})|^{2}\right\} \label{eq:EoM_WW1}
\end{equation}
where the principle value of the first term integral is an integral
avoiding the singularities, one of which resides at $k_{\mathrm{q}}$.

Since the matrix elements of $S_{+}$ and $S_{-}$ are independent
of $k$ and $\chi_{l}(k)$ is an entire function of $k$, $\mathfrak{s}(k)$
in Eq.~(\ref{eq:frak_s}) will not contribute any singularity. The
only other source of singularity is the interaction coefficient $g_{k}$.
For atomic systems in semiclassical treatments for Wigner-Weisskopf
approximation, $g_{k}$ can be regarded as a constant~\cite{Cohen}.
But here, since we treat the spontaneous radiation field as a quantum
field in Eq.~(\ref{eq:rad_Ham}), we adopt the coupling coefficient
to be ($\hbar=1$)~\cite{Mandel} 
\begin{equation}
g_{k}^{2}=\frac{\omega_{\mathrm{q}}^{2}\mu^{2}}{2\epsilon\omega_{k}V}=\frac{ck_{\mathrm{q}}^{2}\mu^{2}}{2\epsilon kV},
\end{equation}
where, for the circuit QED system, $\mu$ denotes the dipole moment
of each qubit and $\epsilon$ the dielectric constant of the waveguide.
$V$ originally designates the volume of the cavity photon, roughly
the box size of the optical cavity. For a superconducting circuit,
$V$ corresponds to the volume of the stripline resonator.

We see $g_{k}^{2}$ is an inverse function of $k$ and it contributes
the other singularity $k=0$ to Eq.~(\ref{eq:EoM_WW1}). Extending
the real variable $k$ to the complex plane, we are able to compute
the integral by replacing the principal value of an integral over
the real line with a contour integral over a closed loop. Since the
singularities only lie on the real line, the contour integral vanish
except for the path along a small semicircle above the singularities.
See App.~\ref{sec:convergence} for the derivation and the proof
that the integral converges. Finally, the equation of motion becomes
\begin{equation}
\frac{\mathrm{d}}{\mathrm{d}t}\alpha(t)=-\alpha(t)\frac{\lambda_{\mathrm{C}}k_{\mathrm{q}}\mu^{2}}{4\epsilon V}\left\{ 2\left|\mathfrak{s}(k_{\mathrm{q}})\right|^{2}-\left|\mathfrak{s}(0)\right|^{2}\right\} \label{eq:EoM_decay}
\end{equation}

\subsection{Decay rate}

From Eq.~(\ref{eq:EoM_decay}), we derive the decay rate
\begin{equation}
\gamma=\frac{k_{\mathrm{q}}\mu^{2}}{4\epsilon A}\left\{ 2\left|\mathfrak{s}(k_{\mathrm{q}})\right|^{2}-\left|\mathfrak{s}(0)\right|^{2}\right\} \label{eq:decay_rate}
\end{equation}
where $A$ denotes the cross-section area of the stripline resonator.

To interpret the expression of the decay rate, we look at its limiting
form. Note that the factor $\mathfrak{s}(k)$, being a function of
$\chi_{l}(k)$ and thus of $k$, also depends on $\ell$ when $k\neq0$.
If the polariton radiation was occurring on an atomic ensemble, this
relative spacing $\ell$ would approach zero and $\mathfrak{s}(k_{\mathrm{q}})$
is identical to $\mathfrak{s}(0)$, which leaves the brace of Eq.~(\ref{eq:decay_rate})
with only one term $|\mathfrak{s}(0)|^{2}$. Furthermore, for an atomic
ensemble in a dielectric, the atomic number $N$ approaches infinity
and $\chi_{l}(k)=1/2$ since the two cosines with $N$ in their arguments
cancel out at the large-$N$ limit. Consequently, $|\mathfrak{s}(0)|^{2}$
becomes a summation of $[S_{l,+}][S_{l,-}]$ over all $l$. In other
words, the decay rate $\gamma$ falls back to a sum rule summing all
transitions back to the ground state, which is no different from the
usual result we have for atomic radiation using Wigner-Weisskopf approximation.

\begin{figure}
\includegraphics[bb=100bp 220bp 500bp 580bp,clip,width=8.5cm]{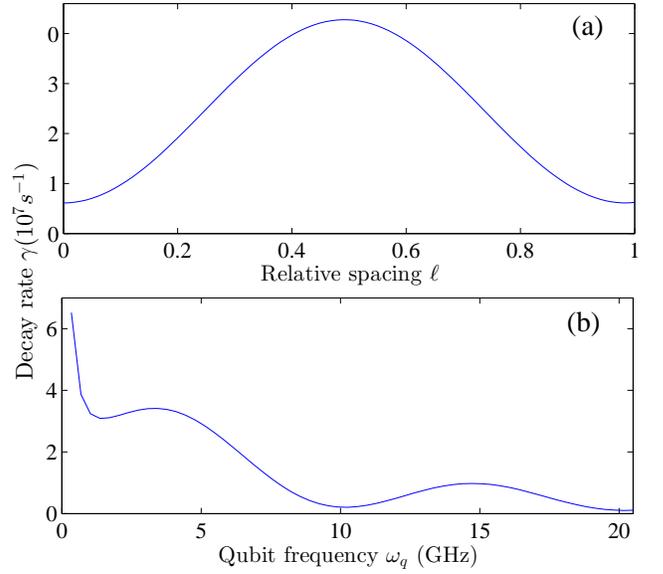}

\protect\caption{Plot of the decay rate of a function of (a) the relative spacing $\ell$
and (b) the qubit frequency $\omega_{\mathrm{q}}$.\label{fig:decay_rate}}
\end{figure}

When we have $\ell\neq0$ for a quasi-lattice of qubits, $|\mathfrak{s}(k_{\mathrm{q}})|\neq\left|\mathfrak{s}(0)\right|$.
Whether this mesoscopic system will have a larger or smaller decay
rate than microscopic atomic ensemble depends on whether $|\mathfrak{s}(k_{\mathrm{q}})|$
is larger or smaller than $|\mathfrak{s}(0)|$. In other words, the
decay is largely determined by the frequency $\omega_{\mathrm{q}}$
and the relative spacing $\ell$ of the qubits in the quasi-lattice.
We illustrate the scenario through the four-qubit ($r=2$) quasi-lattice
case in Fig.~\ref{fig:decay_rate}. The ground state $\left|\mathbf{0},r\right\rangle $
corresponds to $u=-2$ and the excited state $\left|\mathbf{1},r\right\rangle $
corresponds to $u=-1$. The former is composed of a single state $\left|r,u-n;n\right\rangle =\left|2,-2;0\right\rangle $
in the basis before the transformation of Eq.~(\ref{eq:expansion})
and the latter is composed of two states: $\left|2,-2;1\right\rangle $,
where the excitation resides in the cavity field, and $\left|2,-1;0\right\rangle $,
where the excitation resides in the quasi-lattice.

Then according to Eq.~(\ref{eq:matrix_S+}), $S_{+}$ for the ground
to the excited state transition consists of a single entity $2\sqrt{\mathfrak{f}}c_{0}^{(-1)}c_{0}^{(-2)}$.
Using the method outline in App.~\ref{sec:coefficients} for finding
$c_{0}^{(-1)}$ and $c_{0}^{(-2)}$, we have
\begin{equation}
[S_{l,+}]=\frac{4\eta\mathfrak{f}}{\sqrt{\varepsilon^{2}+4\eta^{2}\mathfrak{f}}}\label{eq:S_+_4qubit}
\end{equation}
where $\varepsilon$ is the Stark splitting, separating the substates
in the one-polariton clustered state, i.e. 
\begin{equation}
\varepsilon=\frac{1}{2}\left[\Delta\omega\pm\sqrt{\Delta\omega^{2}+16\eta^{2}\mathfrak{f}}\right].\label{eq:epsilon}
\end{equation}

Using Eq.~(\ref{eq:S_+_4qubit}) for $|\mathfrak{s}(k_{\mathrm{q}})|$
and $|\mathfrak{s}(0)|$, the decay rate as a function of both $\ell$
and $\omega_{\mathrm{q}}$ is plotted in Fig.~\ref{fig:decay_rate}.
The plot in part (a) shows a symmetric decay rate about the qubit
spacing. At the two limiting ends with $\ell=0$ and $\ell=1$ where
the coupling factor $\cos(j\pi\ell)$ falls back to 1, all qubits
in the quasi-lattice are maximally coupled to the cavity field. Under
such a circumstance, each qubit is equally excited and has the least
probability to reabsorb radiated photons, resulting in a minimal probability
of spontaneous radiation and the slowest decay. On the contrary, at
the middle ground with $\ell=1/2$, only half of the qubits (the qubits
at the antinodes) are coupled to the cavity field while the other
half (the qubits at the nodes) are never excited by the cavity photon.
The latter half are all open to reabsorbed the photons emitted from
the former half, maximizing the likelihood of spontaneous emission
across the qubits and giving the largest decay rate for the quasi-lattice.

The inhomogeneous coupling here unevenly excites every other qubit.
When the population of every other qubit is fully inverted, the spin
moment $m$ of the entire quasi-lattice would become zero. Following
Dicke's argument~\cite{dicke54}, this results in a radiated intensity
proportional to $\frac{N}{2}(\frac{N}{2}+1)$, i.e. superradiance.
The prediction of a strong but short radiation pulse matches the largest
decay rate predicted here.

From the plot in Fig.~\ref{fig:decay_rate}(b), we see the decay
also matches with the selective radiation we derived in Sec.~\ref{sec:EoM}A.
At the quasi-period $K$ of Eq.~(\ref{eq:period}) or half of it,
the quasi-lattice are resonant with the cavity where the emitted photons
have the highest probability of being trapped in the cavity and reabsorbed
by the quasi-lattice, giving a minimal decay rate. At the non-resonant
cases, the photons emitting into the waveguide increase and thus the
decay increases.

\section{Conclusion and discussions\label{sec:Conclusion}}

We study the formation of polaritons in a quasi-lattice of superconducting
qubits, that is a linear chain of qubits with inhomogeneous coupling
to a cavity field in a stripline resonator. We show that the radiation
of the quasi-lattice polariton is different from that of an atomic
lattice: the radiation amplitude can be either enhanced or lowered
over the radiation frequency, depending on the resonance between the
polariton and the radiated photon. This amplitude shows a quasi-periodicity
determined by the structure of the quasi-lattice. Further, we find
that the decay of the polariton excited states also depends on the
structure of the quasi-lattice, in particular its spacing $\ell$
between the qubits relative to the cavity wavelength. The decay obtains
its maximum when the qubits in the quasi-lattice are alternatively
excited by the cavity field. These unique features demonstrate the
distinction of the mesocopic nature of superconducting qubit systems
as opposed to the microscopic atoms they emulate.

In addition, since superconducting qubit circuits have played a great
role in the development of quantum computation, the ability to filter
quantum signals selectively using a set of such qubits will benefit
the designs of sophisticated processing devices for quantum signals.
For example, when the qubits are replaced by three-level systems fabricated
on superconducting circuits, sophisticated photon detectors can be
implemented~\cite{romero09,peropadre11}. Further, it was shown coherent
photon transport can be realized on coplanar waveguides with embedded
atoms~\cite{shen05}. Henceforth, if we regard the quasi-lattice
as a mediating device in the coplanar waveguide, complex control over
photon transports by the structure of the quasi-lattice is very likely
to occur.

Overall, the study we present here aims to lay the groundwork for
the physics of many-qubit systems in a circuit cavity. We expect that
more interesting physical phenomena will be discovered when more variables
are added to the system, especially those quantum optical effects
emulatable in a superconducting qubit circuit~\cite{jqyou11}. It
was found that, for example, a single qubit can be dressed by the
cavity field to form an effective three-level system. When a strong
coherent field is added to drive this three-level system, tunable
electromagnetically induced transparency and absorption can be achieved
due to the tunable dressed relaxations~\cite{ian10}. We can resonably
expect that the situation would be much more complex if the quasi-lattice
is driven by the coherent field. For example, while the current experiments
only demonstrate the stimulated emission on a single qubit~\cite{astafiev07,astafiev10}
similar to those in atomic optics, the extension to a quasi-lattice
of qubits might lead to new patterns of stimulated emission. And the
quasi-lattice of qubits is experimentally accessible using current
technologies. 

Another direction worth investigating is to consider that the eigenfrequencies
of the qubits are also non-uniform. In this case, the deformed SU(2)
algebra formed by the quasi-lattice would need further generalizations.
One way to include this non-uniformity into the deformed algebra is
through the statistical approach similar to what we have adopted for
quasi-lattices with non-uniform spacings in Ref.~\cite{ian13}, where
the variations are measured by a pair of mean variance. These two
parameters can then be included in the deformation factor that describes
the quasi-lattice. We hope to extend this consideration and give a
more detailed study in a future paper.
\begin{acknowledgments}
H.I. thanks the support of FDCT of Macau under grant 013/2013/A1 and
University of Macau under grant MRG022/IH/2013/FST. Y.X.L. is supported
by the National Natural Science Foundation of China under Grant Nos.
61025022, 91321208, the National Basic Research Program of China Grant
No. 2014CB921401.
\end{acknowledgments}
\appendix

\section{Matrix elements of the excitation operator\label{sec:Deriving-excitation-operator}}

To verify that the diagonal elements of $S_{+}$ in the polariton
basis vanish, consider for arbitrary $u$
\begin{eqnarray*}
[S_{+}]_{uu} & = & \left\langle u,r\right|S_{+}\left|u,r\right\rangle \\
 & = & \sum_{m,n}c_{m}^{\ast}c_{n}\left\langle m;r,u-m\right|S_{+}\left|n;r,u-n\right\rangle \\
 & = & \sum_{m,n}c_{m}^{\ast}c_{n}\sqrt{\mathfrak{f}(r-u+n)(r+u-n+1)}\\
 &  & \times\left\langle m;r,u-m|n;u-n+1\right\rangle 
\end{eqnarray*}
where the inner product in the last line equals to the Kronecker product
$\delta_{mn}\delta_{u-m,u-n+1}$. Since $m$ cannot simultaneously
equal to both $n$ and $n+1$, the diagonal elements $[S_{+}]_{uu}=0$
for any $u$. Similar arguments also apply to the conjugate $S_{-}$.

For the non-diagonal elements with $u\neq v$, we expand the bra's
and ket's using Eq.~(\ref{eq:expansion}) to get
\begin{align*}
[S_{+}]_{uv}= & \left\langle u,r\right|S_{+}\left|v,r\right\rangle \\
= & \sum_{m,n}c_{m}^{(u)\ast}c_{n}^{(v)}\left\langle r,u-m;m\right|S_{+}\left|r,v-n;n\right\rangle \\
= & \sum_{m,n}c_{m}^{(u)\ast}c_{n}^{(v)}\sqrt{\mathfrak{f}(r-v+n)(r+v-n+1)}\\
 & \times\delta_{m,n}\delta_{u-m,v-n+1}\\
= & \sum_{n}c_{n}^{(u)\ast}c_{n}^{(v)}\sqrt{\mathfrak{f}(r-v+n)(r+v-n+1)}\delta_{u,v+1}.
\end{align*}
The last line in the expansion demands that except for the first off-diagonal,
all other off-diagonal elements vanish, thus giving the expression
in Eq.~(\ref{eq:matrix_S+}). Following the same logic, the lowering
ladder operator has also only the first off-diagonal elements.
\begin{alignat*}{1}
[S_{-}]_{uv}= & \left\langle u,r\right|S_{-}\left|v,r\right\rangle \\
= & \sum_{m,n}c_{m}^{(u)\ast}c_{n}^{(v)}\left\langle r,u-m;m\right|S_{-}\left|r,v-n;n\right\rangle \\
= & \sum_{m,n}c_{m}^{(u)\ast}c_{n}^{(v)}\sqrt{\mathfrak{f}(r+v-n)(r-v+n+1)}\\
 & \times\delta_{m,n}\delta_{u-m,v-n-1}\\
= & \sum_{n}c_{n}^{(u)\ast}c_{n}^{(v)}\sqrt{\mathfrak{f}(r+v-n)(r-v+n+1)}\delta_{u,v-1},
\end{alignat*}
giving the expression of Eq.~(\ref{eq:matrix_S-}).

\section{Commutation relations of the deformed operators\label{sec:comm_relations}}

For the operators $\{S_{z},S_{+},S_{-}\}$ of the collective excitations
on the quasi-lattice, their structure of deformed SU(2) algebra breaks
one of the commutation relation: $[S_{+},S_{-}]\neq2S_{z}$. However,
the other two commutation relations are preserved:
\[
[S_{z},S_{\pm}]=\pm S_{\pm}.
\]
See App.~A of Ref.~\cite{ian12} for a detailed derivation and discussion.

In the transformed polariton basis, $S_{+}$ is expressed using Eq.~(\ref{eq:S+_op})
and Eq.~(\ref{eq:matrix_S+}). We can see the commutation relation
becomes
\begin{align*}
 & \left[\sum_{u}\Omega_{u}\left|u,r\right\rangle \left\langle u,r\right|,S_{+}\right]\\
= & \sum_{u,v}\Omega_{u}[S_{+}]_{v,v-1}\left[\left|u,r\right\rangle \left\langle v-1,r\right|\delta_{u,v}-\left|v,r\right\rangle \left\langle u,r\right|\delta_{v-1,u}\right]\\
= & \sum_{u}\Omega_{u}\biggl\{[S_{+}]_{u,u-1}\left|u,r\right\rangle \left\langle u-1,r\right|\\
 & -[S_{+}]_{u+1,u}\left|u+1,r\right\rangle \left\langle u,r\right|\biggr\}\\
= & \sum_{u}(\Omega_{u}-\Omega_{u-1})[S_{+}]_{u,u-1}\left|u,r\right\rangle \left\langle u-1,r\right|\\
= & \omega_{\mathrm{q}}S_{+}+\sum_{u}(\varepsilon_{u}-\varepsilon_{u-1})[S_{+}]_{u,u-1}\left|u,r\right\rangle \left\langle u-1,r\right|,
\end{align*}
where we have broken up each polariton eigenfrequency $\Omega_{u}$
into two parts: the excitation energy part $u\omega_{\mathrm{q}}$
and the lattice-photon interaction part $\varepsilon_{u}$~\cite{ian13}.
The latter accounts for fine splittings due to the interaction of
each cluster energy level $u\omega_{\mathrm{q}}$. It is determined
by the coupling strength $\eta$ and the detuning $\Delta\omega$.
For low excitation number $u$, its value is less affected by the
number $u$ and the difference between two consecutive ones is minimal
as compared to $\omega_{\mathrm{q}}$, i.e. $\varepsilon_{u}-\varepsilon_{u-1}\ll\omega_{\mathrm{q}}$.

Therefore, we can consider the commutation relation for $S_{+}$ still
preserves after the transformation to the polariton basis. Normal
Baker-Hausdorff formula can then be applied to obtain the Hamiltonian
in the interaction picture as in Eq.~(\ref{eq:Ham_int-pic}).

\section{Convergence in Wigner-Weisskopf approximation\label{sec:convergence}}

First, substituting the expression of $g_{k}^{2}$ into Eq.~(\ref{eq:EoM_WW1}),
we have the equation of motion
\begin{equation}
\frac{\mathrm{d}}{\mathrm{d}t}\alpha(t)=\alpha(t)\frac{\lambda_{\mathrm{C}}k_{\mathrm{q}}^{2}\mu^{2}}{4\pi\epsilon V}\left[\mathscr{P}\negthickspace\int\negthickspace\mathrm{d}k\frac{|\mathfrak{s}(k)|^{2}}{ik(k-k_{\mathrm{q}})}-\frac{\pi}{k_{\mathrm{q}}}|\mathfrak{s}(k_{\mathrm{q}})|^{2}\right].\label{eq:EoM_alpha-1}
\end{equation}
When extending the integration variable $k$ to the complex plane
of variable $z$, the principal value avoids the singularity at $z=0$,
effectively setting the integral as the difference of two integrals
\begin{equation}
\mathscr{P}\int\mathrm{d}z=\lim_{\epsilon\to0}\left[\int_{-\infty}^{-\epsilon}+\int_{\epsilon}^{\infty}\right]\mathrm{d}z=\oint\mathrm{d}z-\int_{\mathrm{sc}}\mathrm{d}z\label{eq:prin_value}
\end{equation}
where $\mathrm{\oint}$ indicates a closed contour integral with an
infinite-radius arc in the upper complex plane and $\int_{\mathrm{sc}}$
indicates a path integral along a small semi-circle over $z=0$.

To verify the path along the infinite-radius arc does not contribute
to the integration, we can first decompose the fraction to have
\begin{equation}
\int\mathrm{d}k\frac{|\mathfrak{s}(k)|^{2}}{ik(k-k_{\mathrm{q}})}=\frac{i}{k_{\mathrm{q}}}\int\mathrm{d}k\left[\frac{|\mathfrak{s}(k)|^{2}}{k}-\frac{|\mathfrak{s}(k)|^{2}}{k-k_{\mathrm{q}}}\right]\label{eq:frac_decomp}
\end{equation}
where $|\mathfrak{s}(k)|^{2}$ is a sum of $|\chi_{l}(k)|^{2}$. It
is not necessary to prove the convergence of the integral from $\chi_{l}(k)$
directly. Since $\chi_{l}(k)$ is a finite sum of the exponential
$e^{ikr_{j}}$, we can simply verify that, in the modulus $|\chi_{l}(k)|^{2}$,
each product term $e^{ik(r_{j}-r_{j'})}$ be convergent with the integration
along the infinite-radius arc. That is,
\begin{eqnarray*}
\int\mathrm{d}k\frac{e^{ik(j-j')L_{\mathrm{q}}}}{k} & = & \int_{\mathrm{arc}}\mathrm{d}z\frac{e^{iz}}{z}\\
 & = & \lim_{|z|\to\infty}\int_{\mathrm{0}}^{\pi}\mathrm{d}\theta|z|ie^{i\theta}\frac{\exp\{i|z|e^{i\theta}\}}{|z|e^{i\theta}}\\
 & = & \lim_{|z|\to\infty}i\int_{\mathrm{0}}^{\pi}\mathrm{d}\theta\left[e^{i|z|\cos\theta}e^{-|z|\sin\theta}\right]
\end{eqnarray*}
For H\"older's inequality, we see the modulus is upper-bounded
\begin{eqnarray*}
\left|\int_{\mathrm{arc}}\right| & \leq & \lim_{|z|\to\infty}\int_{\mathrm{0}}^{\pi}\mathrm{d}\theta\left|e^{i|z|\cos\theta}\right|\cdot\left|e^{-|z|\sin\theta}\right|\\
 & = & \lim_{|z|\to\infty}\int_{\mathrm{0}}^{\pi}\mathrm{d}\theta\left|e^{-|z|\sin\theta}\right|\\
 & = & \lim_{|z|\to\infty}2\int_{\mathrm{0}}^{\pi/2}\mathrm{d}\theta e^{-|z|\sin\theta}.
\end{eqnarray*}

Furthermore, the integral vanishes when the limit is taken. Observe
that for $\theta\in(0,\pi/2)$, the exponential function above is
monotonically non-decreasing. Since the sine function in the first
quadrant is always greater than the diagonal line, i.e. $|z|\sin\theta>|z|\frac{\theta}{\pi/2}$,
by exponentiating both sides, we have
\[
e^{-|z|\sin\theta}<e^{-2|z|\theta/\pi}.
\]
The right hand side can be easily integrated such that
\begin{eqnarray*}
\left|\int_{\mathrm{arc}}\right| & < & \lim_{|z|\to\infty}2\int_{0}^{\pi/2}e^{-2|z|\theta/\pi}\\
 & = & \lim_{|z|\to\infty}\frac{\pi}{|z|}(1-e^{-|z|}).
\end{eqnarray*}
When the limit is taken, we see the path integral vanishes.

For the second term in Eq.~(\ref{eq:frac_decomp}), we see it is
identical to the first term up to an exponential factor:
\[
\int\mathrm{d}k\frac{e^{ik(j-j')L_{\mathrm{q}}}}{k-k_{\mathrm{q}}}=e^{ik_{\mathrm{q}}(j-j')L_{\mathrm{q}}}\int_{\mathrm{arc}}\mathrm{d}z\frac{e^{iz}}{z}
\]
where $z=k-k_{\mathrm{q}}$, i.e. $k$ is extended to the complex
with a translated origin at $k=k_{\mathrm{q}}$. Similarly, it will
also vanish at $|z|\to\infty$. The proof that the contour integral
in Eq.~(\ref{eq:prin_value}) vanishes is now completed and the only
contribution to the principal value is the second integral at $|z|\to0$.

For this integral along a small semicircle above the singularities,
we compute the contribution by each exponential in $|\mathfrak{s}(k)|^{2}$.
For the first term in Eq.~(\ref{eq:frac_decomp}),
\[
\int_{\mathrm{sc}}\mathrm{d}z\frac{e^{iz}}{z}=\lim_{|z|\to0}i\int_{\mathrm{\pi}}^{0}\mathrm{d}\theta\left[e^{i|z|\cos\theta}e^{-|z|\sin\theta}\right]=-i\pi.
\]
The second term has the identical result with the same extra exponential
factor as above, hence
\[
\int\mathrm{d}k\frac{e^{ik(r_{j}-r_{j'})}}{ik(k-k_{\mathrm{q}})}=\frac{\pi}{k_{\mathrm{q}}}\left[1-e^{ik_{\mathrm{q}}(j-j')L_{\mathrm{q}}}\right]
\]

For the total contribution of $|\mathfrak{s}(k)|^{2}$, we then have
\begin{align*}
 & \mathscr{P}\int\mathrm{d}k\frac{|\mathfrak{s}(k)|^{2}}{ik(k-k_{\mathrm{q}})}\\
= & \frac{\pi}{N^{2}k_{\mathrm{q}}}\sum_{Nl,Nl'}^{N-1}\sum_{j,j'}^{N-1}\cos(j\pi l)\cos(j\pi l')\left[1-e^{ik_{\mathrm{q}}(j-j')L_{\mathrm{q}}}\right]\\
 & [S_{l,+}]_{10}[S_{l',-}]_{01}\\
= & \frac{\pi}{N^{2}k_{\mathrm{q}}}\sum_{Nl,Nl'}^{N-1}\left[\chi_{l}(0)\chi_{l'}(0)-\chi_{l}(k_{\mathrm{q}})\chi_{l'}^{\ast}(k_{\mathrm{q}})\right][S_{l,+}]_{10}[S_{l',-}]_{01}\\
= & \frac{\pi}{k_{\mathrm{q}}}\left\{ |\mathfrak{s}(0)|^{2}-|\mathfrak{s}(k_{\mathrm{q}})|^{2}\right\} 
\end{align*}
Plugging the above result into Eq.~(\ref{eq:EoM_alpha-1}), we see
the terms about $k_{\mathrm{q}}$ will cancel out and arrive at the
equation of motion~(\ref{eq:EoM_decay}).

\section{Deriving the coefficients for one-excitation in $N=4$ quasi-lattice\label{sec:coefficients}}

For a $N=4$ quasi-lattice, we have the total spin $r=2$ and the
magnetic moment $m\in\{-2,-1,0,1,2\}$. Since we confine ourselves
to the discussion of the ground and the first excited state, i.e.
the excitation number $u$ being $-2$ and $-1$, respectively, there
are three possible combinations of $n$ and $m$ that satisfies $n+m=u$.
For the ground state $u=-2$, we have one configuration in the expansion
of Eq.~(\ref{eq:expansion}), so
\[
c_{0}^{(-2)}=1.
\]

For the coefficients in the expansion of the first excited polariton
state, we can either plugging in the numbers into Eq.~(\ref{eq:coeff})
or follow the routine of finding a set of difference equations~\cite{ian12}.
In this case, the latter is simpler and we have
\begin{eqnarray*}
c_{1}^{(-1)} & = & \frac{\varepsilon}{2\eta\sqrt{\mathfrak{f}}}c_{0}^{(-1)},\\
0 & = & \frac{\varepsilon-\Delta\omega}{\eta\sqrt{2\mathfrak{f}}}c_{1}^{(-1)}-\sqrt{2}c_{0}^{(-1)}.
\end{eqnarray*}
Combining these two equations, we find a quadratic equation for $\varepsilon$
\[
\varepsilon^{2}-\Delta\omega\varepsilon-4\eta^{2}\mathfrak{f}=0,
\]
the solution of which is given in Eq.~(\ref{eq:epsilon}).

Further, using the normality condition for the superposition coefficients,
we get
\begin{eqnarray*}
c_{0}^{(-1)} & = & \sqrt{\frac{4\eta^{2}\mathfrak{f}}{\varepsilon^{2}+4\eta^{2}\mathfrak{f}}},\\
c_{1}^{(-1)} & = & \sqrt{\frac{\varepsilon^{2}}{\varepsilon^{2}+4\eta^{2}\mathfrak{f}}}.
\end{eqnarray*}
Then, substituting the coefficients into the expression $2\sqrt{\mathfrak{f}}c_{0}^{(-1)}c_{0}^{(-2)}$
for $[S_{l,+}]$, we arrive at Eq.~(\ref{eq:S_+_4qubit}).

\end{document}